\documentclass[11pt,twoside]{article}
\usepackage{asp2010,natbib,sidecap,graphicx}

\resetcounters

\begin{document}

\title{X-ray sources in Galactic old open star clusters}
\author{Maureen van den Berg$^{1,2}$
\affil{$^1$Astronomical Institute Anton Pannekoek,
  University of Amsterdam,\\ Science Park 904, 1098 XH Amsterdam, The
  Netherlands; m.c.vandenberg@uva.nl\\
  $^2$Harvard-Smithsonian Center for Astrophysics,
    60 Garden Street, Cambridge, 02138 MA, USA}
}

\begin{abstract}
I review the current status of studies of the X-ray sources in
Galactic old open clusters. Cataclysmic variables (CVs),
magnetically-active binaries (ABs), and sub-subgiants (SSGs) dominate
the X-ray emission of old open clusters. Surprisingly, the number of
ABs detected inside the half-mass radius with
$L_X\gtrsim1\times10^{30}$ erg s$^{-1}$ (0.3--7 keV) does not appear
to scale with cluster mass. Comparison of the numbers of CVs, ABs, and
SSGs per unit mass in old open and globular clusters shows that each
of these classes is under-abundant in globulars. This suggests that
dense environments suppress the frequency of even some of the hardest
binaries.
\end{abstract}

\section{Introduction}

The fate of star clusters and the binaries in them are closely
intertwined. Dynamical encounters between single stars and binaries
affect the stellar velocity distribution, and thus the evolution of
the cluster as a whole; vice versa, the dense environment of a cluster
can trigger interactions that affect the binary properties. Open
clusters allow detailed studies of complete samples of binaries, with
the advantage that their age, distance, and composition are known from
cluster membership. Surveys like the WIYN Open Cluster Study
\citep{math00} now enable a detailed comparison between observations
and numerical models of the evolution of stellar clusters and their
binary populations \citep[e.g.][]{hurlea05}. Until the early 1990s,
the study of old open clusters was not pursued with X-ray
observations. The reason for this is that the expected contribution of
single stars to the X-ray emission of old stellar populations is
small. In low-mass stars X-rays are emitted by the hot gas of the
corona surrounding the star. The level of X-ray emission is linked to
the stellar rotation rate, as rotation powers the magnetic dynamo that
is responsible for heating the corona; the lower the spin rate, the
weaker the X-ray emission. The rotation of late-type single stars
declines over time as a result of angular-momentum loss through wind
outflows along the stars' magnetic field.

For stars that are part of a binary, a high level of X-ray emission
can be maintained despite an advanced age. Systems in which mass is
exchanged between the binary companions are bright in X-rays if one of
the stars is a compact object. Examples are cataclysmic variables
(CVs) where the compact star is a white dwarf, or low-mass X-ray
binaries (LMXBs) that contain a neutron star or black hole. Strong
tidal interaction between detached main-sequence stars in a close
binary locks their rotation to the orbital period. The rapid rotation
induced by the synchronization keeps the dynamo active, and thus the
process responsible for the X-ray production. Such systems are called
magnetically-active binaries, or ABs in short. For a study of the
interacting binaries in an old stellar population, X-ray emission is
an excellent tracer.

X-ray observations are therefore crucial for studying how the
environments of star clusters affect the binary numbers and binary
properties. In this paper I review what we have learned from X-ray
studies of old open clusters, a field to which Utrecht has contributed
significantly. I start with an overview of X-ray studies of old open
clusters in Sect.~\ref{sec_obs}.  In Sect.~\ref{sec_classes} I
describe the various classes of open-cluster X-ray
sources. Sect.~\ref{sec_glob} compares the X-ray properties of old
open and globular clusters. The chapter by Frank Verbunt gives an
overview of X-ray studies of globular clusters.

\section{An overview of X-ray observations of old open clusters} \label{sec_obs}

\subsection{{\em ROSAT}} \label{sec_rosat}

The first old open cluster that was the target of a pointed X-ray
observation is M\,67 (NGC\,2682). Belloni, Verbunt, and collaborators
used the {\em ROSAT} Position Sensitive Proportional Counter (PSPC) to
observe this 4-Gyr old cluster down to a limit of about $2 \times
10^{30}$ erg s$^{-1}$ (0.1--2.4 keV) in 1991.  The photometric
monitoring campaign by \cite{gillea}, aimed at looking for solar-like
oscillations in members of M\,67, had just serendipitously discovered
the first CV in an open cluster (EU\,Cnc). The light curve of this
optically-faint object showed large-amplitude brightness variations at
a period of 2.09 h that resembled the variability of so-called AM\,Her
systems---CVs that contain a white dwarf with a strong magnetic field
($B \gtrsim 10$ MG). The main motive for the X-ray observation by
Belloni et al.~was to do X-ray follow-up for this specific object; as
CVs with accurate distance and reddening estimates were (and still
are) rare, this was an excellent opportunity for an accurate
measurement of the intrinsic X-ray luminosity (assuming cluster
membership). The X-ray counterpart was readily detected, and the
softness of its X-ray spectrum in the {\em ROSAT} band agreed with its
AM\,Her classification \citep{bellea93}. Later, the magnetic nature
was also confirmed by its optical spectrum \citep{pasqbellea} and {\em
  ROSAT} light curve \citep{bellea}.

Although this initial observation of M\,67 was not optimally pointed
at the cluster center, it did reveal that at least six more likely
members of M\,67 were similarly bright in the {\em ROSAT} band as the
CV. Thanks to the wealth of optical information available for this
well-studied cluster, it quickly became clear that most of these (in
fact: {\em all} of these, as we now know) are binaries with periods
$\lesssim$45 d, including many with circular orbits. Since the theory
of tidal interaction predicts that synchronization occurs before
circularization \citep[e.g.\,][]{zahn89}, this suggests that in these
systems the stellar rotation is coupled to the orbit. \cite{bellea93}
therefore suggested that these sources are ABs in which the X-rays are
the result of magnetic activity. A second {\em ROSAT} observation
(this time centered on the cluster), combined with new optical
results, detected many other binaries in M\,67 and refined the
classification of some of the earlier detections
\citep{bellea}. Surprisingly, this old cluster that at first seemed
like an unexciting target for an X-ray study, revealed an enormous
variety of X-ray sources. More details on the different source classes
are given in Sect.~\ref{sec_classes}.

\begin{SCfigure}[2.5][h]
\includegraphics[width=4.9cm]{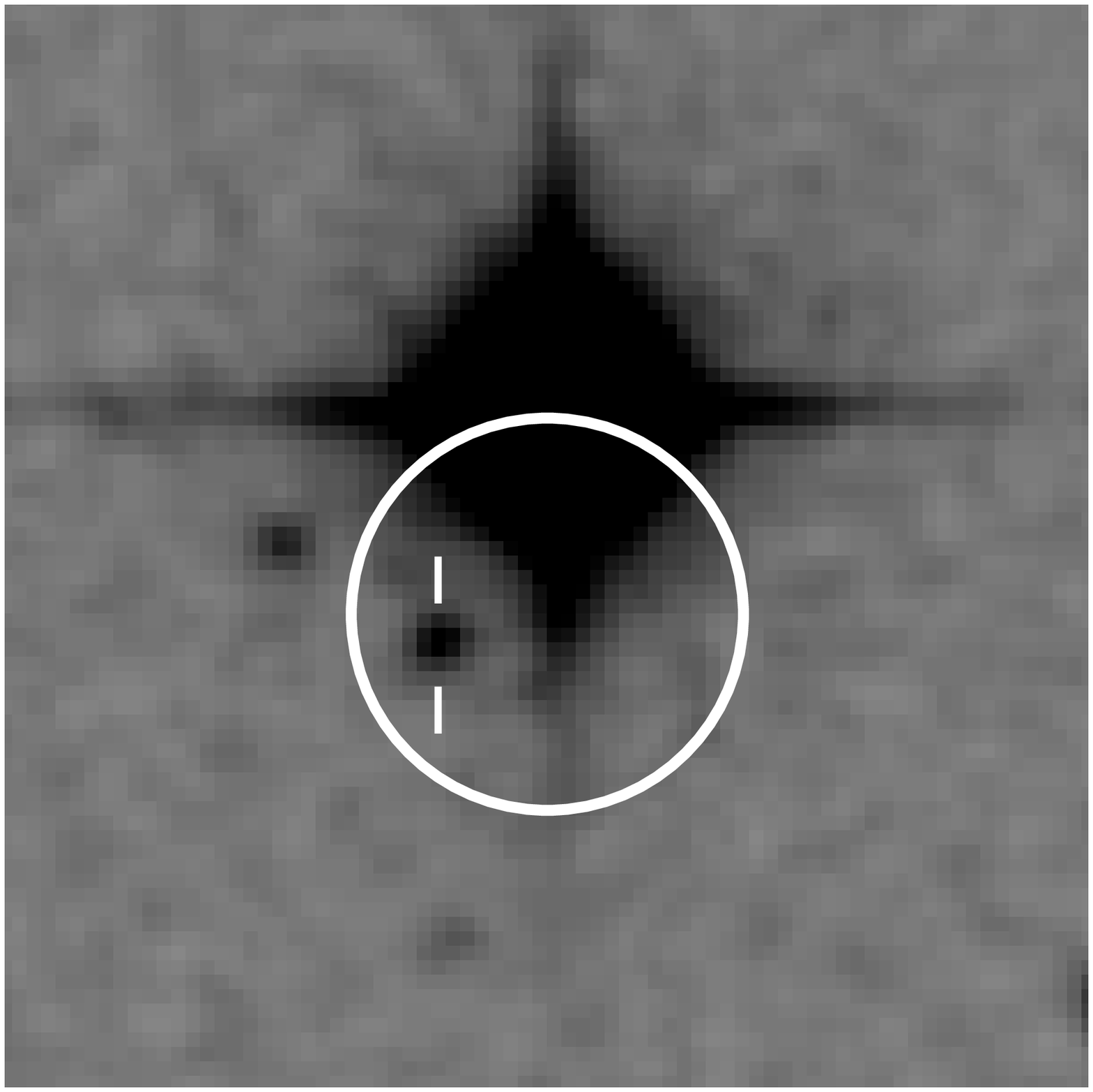} \vspace{0.5cm}
\caption{{\small DSS image (1.25\arcmin\,$\times$ 1.25\arcmin\,)
    around the source X\,45 in NGC\,752, illustrating the possibility
    for spurious identifications of {\em ROSAT} sources. The circle
    marks the 90\% error radius. \cite{bellverb} suggested that the
    wide binary and blue straggler H\,209 (bright star) is the
    counterpart, but optical follow-up by \cite{vdberg2001c} found no
    clues to the origin of the X-rays. An alternative $V\approx18.9$
    counterpart (vertical tick marks) is included in the GSC 2.3
    catalog, but not in the GSC 1.0 catalog used for the optical
    identification.}}
\label{fig_h209}
\end{SCfigure}

Several other old open clusters, both younger and older than M\,67,
were studied with {\em ROSAT}. In order of increasing age, these are
NGC\,6940 \citep[0.6--1 Gyr; ][]{belltagl}, IC\,4651 \citep[1--1.5
  Gyr; ][]{belltagl98}, NGC\,752 \citep[2 Gyr; ][]{bellverb}, and
NGC\,188 \citep[6.5 Gyr; ][]{bellea}. Classification of the X-ray
sources in each of these resulted in a similar picture as for M\,67:
cross-correlation of the source lists with optical catalogs pointed at
a high incidence of binaries among the candidate counterparts. An
unexpected discovery was that some known binaries that were detected
did not fit the profile of being circular, tidally-locked ABs. For
example, some {\em ROSAT} sources were identified with eccentric
systems having orbital periods of years, far too long for tidal
coupling strong enough to induce rapid rotation. At present we still
do not understand the X-rays of most of these long-period binaries
(Sect.~\ref{sec_longperiod}). Overviews of the {\em ROSAT} results can
be found in \cite{bell97} and \cite{verb00}. None of the other
clusters showed an X-ray source population so rich and varied as
M\,67. However, as M\,67 was studied with the highest sensitivity (in
part because of its proximity and low reddening) or with the largest
fractional coverage, a systematic comparison between clusters based on
the {\em ROSAT} data is difficult to make.

Positional uncertainties of the {\em ROSAT} sources range from about
5\arcsec\, to 25\arcsec, and in some cases multiple candidate
counterparts lie inside the error circles. By repeating their
X-ray/optical matching with artificial source lists, Belloni et
al.~estimated that the chances for random coincidences were relatively
small, and in fact many proposed identifications were later confirmed
with {\em Chandra} or {\em XMM-Newton} observations. Indeed, the
fraction of known binaries among the possible counterparts is too
large to be entirely coincidental. On the other hand, in individual
cases it is wise to be wary of the possibility of a spurious match. An
example is the source X\,45 in NGC\,752, which \cite{bellverb}
identified with the blue straggler and long-period binary H\,209 in
NGC\,752. This identification prompted a comparison with the detection
of the blue straggler S\,1082 in M\,67, which was found to be a
(possibly physically-bound) multiple system consisting of a
long-period and short-period binary, in which the latter is
responsible for the X-rays (see also Sect.~\ref{sec_longperiod}). But
extensive optical follow-up by \cite{vdberg2001c} did not reveal any
sign of a close binary in H\,209. While it cannot be excluded that the
parameters of H\,209 are unfavorable for finding optical signatures
for S\,1082-like multiplicity, the option that the fainter optical
source in the 90\% error circle is the true counterpart should be
seriously considered (Fig.~\ref{fig_h209}). Low-resolution optical
spectra are needed to classify this alternative counterpart; given the
$\sim$17\arcsec-separation from the bright blue straggler, such
spectra should be easy to acquire.

\subsection{{\em Chandra} and {\em XMM-Newton}}

With {\em Chandra} and {\em XMM-Newton} more sensitive studies have
been done of clusters that had already been observed with {\em
  ROSAT}. \cite{vdbergea04} describe a detailed {\em Chandra} study of
M\,67, NGC\,188 was observed with {\em XMM-Newton} \citep{gond05}, and
a deep study of NGC\,752 with both satellites---mainly aimed at
studying the X-ray emission of single solar-type stars---can be found
in \cite{giarea08}. The sensitivity and positional accuracy of the new
instruments also allowed more distant, compact, or reddened open
clusters to be studied, thus enabling exploration of the X-ray source
populations over a broader range of cluster parameter space.  So far,
especially the addition of the old (8 Gyr), massive cluster NGC\,6791
\citep{vdbergea12ba} has been useful for gaining new insights thanks
to the many X-ray sources detected and the extensive body of available
optical data (deep photometry, proper motions, variability, follow-up
spectroscopy). There are now two parallel ongoing efforts aimed at
studying the X-ray properties of old and intermediate-age open
clusters. The {\em Chandra} survey by van den Berg et al.~focuses on
the oldest clusters ($\gtrsim 3$ Gyr) while the survey by Pooley et
al.~with {\em XMM-Newton} also includes several younger ones.

{\em Chandra} and {\em XMM-Newton} observations have not (yet) led to
the discovery of any fundamentally different source classes than those
already found with {\em ROSAT}, although, if validated by follow-up
spectroscopy, the candidate quiescent low-mass X-ray binary (qLMXB) in
NGC\,6819 reported by \cite{gosnea12} could turn out to be the first
of its kind to be uncovered in an open cluster
(Sect.~\ref{sec_other}). However, the much-improved positional
accuracy has significantly reduced the chances for spurious
identifications, and the broader spectral response has facilitated
source classification.

\section{X-ray source classes} \label{sec_classes}

\subsection{Cataclysmic variables}

After the discovery of EU\,Cnc in M\,67, four more open-cluster CVs
were found, all in NGC\,6791 \citep{kaluea97,vdbergea12ba}. EU\,Cnc is
quite faint ($\sim$4$\times$10$^{29}$ erg s$^{-1}$, 0.3--7 keV), while
B\,8 in NGC\,6791 is two orders of magnitude brighter. Most were first
identified as CV candidates through their optical colors or
variability, while CX\,19 in NGC\,6791 is the first X-ray--selected CV
in an open cluster. Its proposed optical counterpart was chosen as a
high-priority target for follow-up spectroscopy based on the blue
color; Balmer and He~II emission lines confirm its CV nature. For CVs
in globular clusters it is a point of discussion as to whether they
have intrinsically different properties than those in the field. The
on-average larger X-ray luminosities, higher X-ray--to--optical flux
ratios, and possibly lower outburst rates for globular-cluster CVs
could perhaps be explained if there is a prevalence of magnetic
systems among them, or if the typical accretion rates are lower
\citep{edmogillea03b}. These differences could then be related to
their formation scenarios, which may have involved dynamical
interactions. As there are only a handful of confirmed CVs in open
clusters, it is difficult to make a comparison with either the field
or globular-cluster CV population. Besides EU\,Cnc, CX\,19 is possibly
magnetic \citep{vdbergea12ba}, which is in line with the finding that
many field CVs that turn out to be magnetic are bright in X-rays and
are first discovered in X-rays. What can be said is that scaling the
number of CVs in NGC\,6791 by cluster mass results in a CV density
that is consistent with estimates for the CV density in the field,
which points at a primordial origin and does not require any
dynamical formation or destruction processes.

If the density of CVs in open clusters is indeed similar as in the
field, the reason why some CV searches have not been successful is
likely because of the relatively small number of cluster members
surveyed \citep[e.g.\,][]{kafkea04}. A few candidate CVs in open
clusters have been identified through dwarf-nova--like outbursts
(V\,57 in in the 2--3 Gyr old NGC\,2158, \cite{mochea06}; 15877\_2 in
the $\sim$3.5 Gyr old NGC\,6253 \cite{demaea10}), or their X-ray
spectral properties and UV/optical colors (X\,2 in NGC\,6819,
\citealt{gosnea12}). More follow-up is needed to establish the nature
and membership status of these candidates. Some of the faint, blue
candidate counterparts to {\em Chandra} or {\em XMM-Newton} sources
may also turn out to be CVs (or other compact accreting binaries),
although, at least for those in M\,67 and NGC\,6791, it is estimated
that they are mainly background galaxies.

\subsection{Active binaries} \label{sec_abs}

Active binaries, including both detached systems and contact binaries,
are the most common open-cluster X-ray sources. They are among the
brightest sources, but found down to the detection limit of the
observations ($\sim$2$\times$10$^{28}$ erg s$^{-1}$, 0.3--7 keV, for
M\,67). The study of coronal activity of M\,67 sources by
\cite{pasqbell} found no relation between the {\em ROSAT} X-ray
luminosity and stellar parameters like optical magnitude or orbital
period. As we now know, this was the result of wrong or incomplete
information on orbital or rotation periods, by the inclusion of
sources with likely more complex evolutionary histories than regular
ABs, and by the high limiting flux of the initial {\em ROSAT} pointing
(which, as it turns out, was only sensitive enough to detect regular
ABs with orbital periods between $\sim$0.5 and 1.5 d). Later studies
with larger, and cleaner, AB samples {\em do} reveal an
activity-rotation relationship in the sense that the coronal X-ray
luminosity decreases as a function of orbital period up to the
limiting period for tidal circularization
\citep{vdbergea04,vdbergea12ba}; for M\,67 this period lies around 12
d. The explanation is that the stellar rotation, and hence the level
of activity, is lower in tidally-circularized (and thus synchronized)
binaries with longer periods.  As is seen for the X-ray emission of
single stars \citep[e.g.\,][]{rand97}, saturation of the X-ray
activity occurs for stars in binaries with the shortest orbital
periods and highest rotation rates, such as contact binaries. Since
the time scale for tidal synchronization is shorter than for
circularization, it is possible that the rotation of stars in
eccentric binaries is locked to the orbital period around periastron
where the interaction is strongest (so-called
pseudo-synchronization). If the resulting rotation is fast enough to
generate activity, such binaries can also show up as ABs. An example
is the $\sim$32-d period binary S\,1242 in M\,67, which has an
eccentricity of 0.66 and whose photometric period of 4.88 d
\citep{gillea} corresponds to the corotation period at periastron.

\cite{vdbergea12ba} compared the number of ABs inside the half-mass
radius with $L_X \gtrsim 1\times 10^{30}$ erg s$^{-1}$ (0.3--7 keV)
for three old open clusters that were observed with {\em Chandra} or
{\em XMM-Newton}. Surprisingly, they find that this number does not
scale with cluster mass, as would be expected for a primordial
population. NGC\,6791 is 4.5--6.4 times more massive than M\,67, but
has 0.9--1.6 times the number of ABs. NGC\,6819 has 2.4 times the mass
of M\,67, but only has a fraction ($\sim$0.13) of the number of
ABs. At this point we have no explanation for this, and (deeper)
studies of more clusters are required. Possibly, M\,67 has lost a
higher fraction of its initial mass compared to the other two
clusters. If it lost preferentially single, low-mass stars through
evaporation while retaining more binary stars that sunk to the core
due to mass segregation, the current binary population would appear as
representative of a much more massive cluster. The small number of
X-ray sources in NGC\,188 could therefore not just be the result of
limited sensitivity of the {\em ROSAT} pointing, as was suggested by
\cite{verb00}.

\subsection{Sub-subgiants}

Among the brightest (up to $\sim$10$^{31}$ erg s$^{-1}$, 0.3--7 keV)
X-ray sources in old open clusters are binaries that lie below or to
the red of the sub-giant branch, the so-called sub-subgiants or red
stragglers. This name derives from the fact that their photometry
cannot be explained by the combined light of two ordinary cluster
members. Over a dozen sub-subgiants are known in open and globular
clusters, and they are typically detected in X-rays. All show signs of
binarity, but they include very distinct source classes (detached
binaries, CVs, and at least one neutron star with an evolved
companion). It is possible that they exist in the field as well, but
the poorly constrained ages and distances of field stars make it
difficult to recognize them as sub-subgiants.

Given their X-ray spectral properties and signs of chromospheric
activity (Ca\,II H\&K and H$\alpha$ emission), the X-ray emission of
sub-subgiants in old open clusters is likely the result of coronal
activity. This explanation is not without problems for all systems,
though. While the optical variability of the sub-subgiant S\,1113 in
M\,67 suggests that the rotation is synchronized to the orbit
\citep{vdbergea02}, the light curve of the other M\,67 sub-subgiant
S\,1063 (an eccentric binary in a 18.4-d period) displays what looks
like spot modulation on a period that is longer than both the
pseudo-synchronous and orbital period (Fig.~\ref{fig_s1040}). Tidal
locking, which sets the stellar rotation in normal ABs, may not have
been achieved yet, or another mechanism drives the spin rate in
S\,1063.  A recent or ongoing episode of mass transfer has been
invoked to explain the optical photometry of CVs in 47\,Tuc that lie
in the sub-subgiant region of the color-magnitude diagram
\citep{albrea01}. But for S\,1063 the eccentric orbit argues against
mass transfer in the recent past, as a (nearly) Roche-lobe filling
star would have quickly circularized the orbit \citep{mathea03}.

Sub-subgiants are not uncommon in old open clusters, and their numbers
appear to scale with cluster mass \citep[][see also
  Table~\ref{tab_glob}]{vdbergea12ba}. This suggests that the
explanation of their anomalous properties lies in a hitherto
overlooked binary-evolution path, and is not the result of a
short-lived perturbed state. N-body simulations by \cite{hurlea05}
created one star below the sub-giant branch, which resulted from the
merger of two stars in a binary after instable mass transfer. This
star is single though, and some basic ingredient (primordial triples?)
is still missing from the models.

\subsection{Long-period binaries, and blue and yellow stragglers} \label{sec_longperiod}

Another class of poorly-understood X-ray sources are those that are
identified with wide binaries. In such systems tidal interaction is
far too weak to lead to enhanced stellar rotation rates. In most cases
the suggested optical counterparts also have anomalous optical
properties: based on their photometry they are classified as blue or
yellow stragglers, which is a sign of a complex past that may have
involved mass transfer, a merger, or perhaps some kind of dynamical
encounter. It is not clear if their X-ray emission is always
intrinsically linked to their evolutionary histories. An example of a
wide spectroscopic binary with ``normal'' colors is the {\em ROSAT}
source \#6 (VR\,111) in NGC\,6940, which has a period of almost 3600 d
\citep{belltagl}.

Blue stragglers are found in most old clusters, and it is a matter of
debate if they are formed through binary evolution, dynamical
encounters, or both \citep[see e.g.\,][]{mathgell09}. X-ray emission
is not a common property of blue stragglers. As the X-rays indicate
that some kind of binary interaction is currently active, tracking
down their origin can also provide clues to what led to a system's
formation. The best example is S\,1082 in M\,67. Studies triggered by
its X-ray detection showed that it contains two stars that are blue
stragglers on their own account, and that a close and wide binary
contribute to the optical light \citep{vdbergea2001ad,sandea03}. If
these are bound, at least six stars must have been involved in
creating S\,1082, making a dynamical formation most likely
\citep{leigsill11}.  Other X-ray--emitting blue stragglers in wide
orbits are S\,997 in M\,67 \citep{vdbergea04}, H\,209 in NGC\,752
(Sect.~\ref{sec_rosat}), and possibly L\,44 in IC\,4651
\citep{belltagl98}.

\begin{SCfigure}[2.5]
\includegraphics[width=5.5cm]{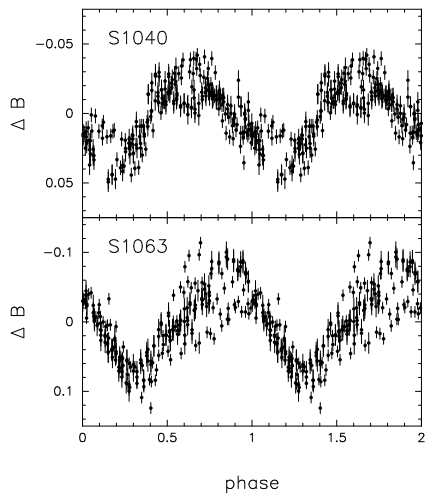}
\caption{{\small YALO light curves from 2001 Oct to 2003 Jun of the
    yellow-straggler and sub-subgiant binaries S\,1040 ({\em top}) and
    S\,1063 ({\em bottom}) in M\,67. For S\,1040 the data are folded
    on a period of 47.05 d, for S\,1063 on 23.38 d. For both,
    photometric variability is likely caused by spots near active
    regions on the surface. The modulation periods are longer than the
    orbital periods and, for S\,1063, also longer than the
    pseudo-synchronous rotation period.  This suggests that the
    rotation is not synchronized, and therefore that it is not (only)
    tidal coupling that is responsible for the enhanced magnetic
    activity. }
\label{fig_s1040}}
\end{SCfigure}

M\,67 features three yellow stragglers, i.e.~stars between the blue
stragglers and the red-giant branch, which all have wide orbits
($\gtrsim$40 d) and secure {\em Chandra} detections. The eccentric
binaries S\,1072 and S\,1237 show no signs of chromospheric activity
\citep{vdbergea}; possibly a faint close binary in these systems is
overwhelmed by the light of the primary. S\,1040 is an interesting
system for which \cite{verbphin} predicted the presence of a
white-dwarf secondary based on the circular, 42.8-d orbit. They argued
that the radius of the yellow-straggler primary is too small to be
responsible for the circularization, which must then be attributed to
a former primary that was much larger in the past and filled its Roche
lobe. The white dwarf was indeed discovered in a UV pointing of M\,67
\citep{landea}. Magnetic activity of the yellow straggler is
manifested in the coronal X-ray emission, but also in chromospheric
emission lines \citep[e.g.\,][]{vdbergea}, and is possibly a relic of
the previous phase of mass transfer. The rotation rate as derived from
photometric variability is lower than expected for synchronous
rotation (see Fig.~\ref{fig_s1040}).

\subsection{Other source classes} \label{sec_other}

Continued loss of angular momentum in a contact binary can lead to a
merger of the two stars. It is expected that the result is a single,
rapidly-rotating star. The brightest ($\sim$2$\times$10$^{31}$ erg
s$^{-1}$, 0.1--2.4 keV) {\em ROSAT} source X\,29 in NGC\,188 was
identified with such an FK\,Com-type giant \citep{bellea}. As the
X-rays are driven by rotation, this class of coronal emitter is
closely related to the ABs.

A few X-rays sources have been identified with cluster giants that
show no signs of binarity. Some are among the brightest sources in the
cluster. Examples are the {\em ROSAT} sources \#13 (VR\,108) in
NGC\,6940 \citep{belltagl} and X\,19 (S\,364) in M\,67 \citep{bellea},
and the {\em Chandra} source CX\,9 in NGC6791
\citep{vdbergea12ba}. These could be truly single stars or very wide
binaries; in both cases the X-rays remain a mystery. Alternatively,
these stars could be spurious matches (although VR\,108 does show weak
signs of Ca~II H\&K activity; \cite{vdberg2001c}).

Optical spectroscopy of the faint blue counterpart to a very soft {\em
  ROSAT} source revealed it to be a hot white dwarf
\citep{pasqbellea}. There are no indications for binarity of this
star, and given the estimated temperature of about 68\,000 K
\citep{flemea97} it is most likely a thermal X-ray emitter. While
Pasquini et al.~argue that it is a member because single
X-ray--emitting white dwarfs are rare, the proper-motion study by
\cite{yadaea08} suggests a low probability for cluster membership.

Based on its soft X-ray spectrum and limits on the X-ray--to--optical
flux ratio, \cite{gosnea12} tentatively classify the brightest X-ray
source detected in the field of NGC\,6819 as a qLMXBs. As the chances
of finding any primordial LMXBs in a cluster the size of NGC\,6819 are
low, they suggest that---if its qLMXB nature is confirmed---this
object likely has a dynamical origin. So far, no qLMXBs have been
found in open clusters. \cite{vdbergea04} suggested that the soft and
highly variable X-ray source CX\,2 in M\,67 could be a good candidate,
but follow-up spectroscopy showed that the candidate optical
counterpart is an active galaxy.

\section{Comparison with globular clusters} \label{sec_glob}

It has been known for a long time that as a result of their high
central stellar densities, globular clusters are very efficient in
forming objects that are rare in the field such as LMXBS and their
descendants, the milli-second pulsars or MSPs (see the contribution by
Frank Verbunt). \cite{verb00} pointed out first that, when bright
($\gtrsim 10^{36}$ erg s$^{-1}$) LMXBs in globular clusters are
disregarded, the integrated {\em ROSAT} X-ray luminosities per unit
mass of most globular clusters is lower than that of M\,67. This
raised the questions of whether globular clusters are efficient at
destroying those binaries that are responsible for most of the X-ray
emission in old open clusters (i.e.~ABs in the case of M\,67), or
whether M\,67 contains an exceptionally high fraction of ABs. At the
time, this problem could not be tackled directly by observations
because of the lack of sensitivity and spatial resolution to detect
and identify ABs in globular clusters.

\begin{table*}\small
\begin{center}
\begin{tabular}{lccccccc}
\hline
\hline
cluster     & age        & $M$              & $N_X$      & $N_{X, CV}$  & $N_{X, S}$ & $N_{X, AB}$ & $\log (2~L_{30}$/$M$) \\
            & (Gyr)      & ($M_{\odot}$)     &            &             &          &            &         \\
\hline
NGC\,6819   &  2--2.4    & 2600             & {\em 6--7} &  {\em 1?}  &  {\em 1?} &   {\em 1?} &   \ldots            \\ 
M\,67       &  4         & 1100             & 12         &    0       & 1         &  7--8      &  28.9        \\ 
NGC\,6791   &  8         & (5--7)$\times$$10^{3}$ & 15--19     &   3--4     & 3         & 7--11      &  28.6--28.8           \\ 
\hline
47\,Tuc     & 11.2       & $1.3\times10^{6}$ & $\sim$200 & 30--119    & 10        & 42--131   & 28.0 \\
NGC\,6397   & 13.9       & $2.5\times10^{5}$ & 15--18    &  11        & 2         & 0--2      & 27.7\\

\hline
\end{tabular}
\caption{X-ray sources in globular and old open clusters inside the
  half-mass radius $r_h$ with $L_X\gtrsim1\times10^{30}$ erg s$^{-1}$
  in the 0.3--7 keV band. I list the number of sources identified with
  cluster members ($N_X$), CVs ($N_{X,CV}$), sub-subgiants
  ($N_{X,S}$), and ABs ($N_{X,AB}$). The numbers for NGC\,6819 are
  uncertain due to limited optical follow up. Log (2 $L_{30}$/$M$) is,
  in log units, the ratio of the total X-ray luminosity of the $N_X$
  sources to the cluster mass ($M$) divided by two to account for the
  selection of sources inside $r_h$. The horizontal line separates
  open and globular clusters. Main references: \cite{gosnea12},
  \cite{vdbergea04, vdbergea12ba}, \cite{cohnea10},
  \cite{heinea05b}. See \cite{vdbergea12ba} for
  details. \label{tab_glob}}
\end{center}
\end{table*}

With the excellent imaging capabilities of {\em Chandra} we can
revisit this issue. Although almost 80 globular clusters have been
studied with {\em Chandra}, only a handful have been observed with
sufficient sensitivity to access a significant part of the AB X-ray
luminosity function. \cite{vdbergea12ba} made a detailed comparison
between the numbers of CVs, ABs, and sub-subgiants down to $L_X = 1
\times 10^{30}$ erg s$^{-1}$ in those old open and globular clusters
for which the most comprehensive source classifications are available
(see also Table~\ref{tab_glob}). It turns out that both suggested
scenarios appear to be relevant for explaining the integrated X-ray
properties of old clusters. As discussed in Sec.~\ref{sec_abs}, among
old open clusters M\,67 has a high AB frequency for its mass. On the
other hand, not just M\,67 but also NGC\,6791 has a higher X-ray
emissivity than the two globular clusters. All three main source
classes are under-represented in globulars when scaling by mass. The
specific frequency of CVs appears to be less disparate than that of
ABs; a possible reason is that in globulars they are dynamically
created as well \citep{poolhut06}. The X-ray emission from qLMXBs and
MSPs in globular clusters, i.e.~types of faint X-ray sources of which
there are no confirmed cases in old open clusters, cannot make up for
the lack of CVs and coronal sources. In a study of a larger sample of
globular clusters, Heinke et al.~(in preparation) find that their
X-ray emissivity is lower than that of old low-density environments in
general.

The suppressed numbers of CVs and ABs is in line with the overall low
binary frequency in globular clusters compared to open clusters
\citep{miloea12}. This suggests that in the intricate process of
forming and breaking up binaries in the cores of globulars, the
balance is tipped in favor of binary destruction---not just for wide
systems but also for close, interacting binaries that dominate the
X-ray emission of old populations. Much remains to be done to
classify, dissect, and compare the populations of faint X-ray sources
in different settings in more detail. Old open clusters, with their
rich X-ray source populations that are relatively accessible for
optical follow-up work, can play a crucial part in these studies.

\acknowledgements I would like to thank Frank Verbunt, Josh Grindlay,
Haldan Cohn, Phyllis Lugger, and Craig Heinke for many discussions on
the topic of X-ray sources and close binaries in old open and globular
clusters. This research was partly supported by various {\em Chandra}
grants for GO programs.

\bibliographystyle{asp2010}
%\bibliography{biball}

\end{document}